\documentclass[aps,prb,showpacs,superscriptaddress,twocolumn]{revtex4}

\usepackage{amsmath,amssymb}
\usepackage{graphicx}
\usepackage{longtable}
\usepackage[ansinew]{inputenc}
\usepackage{color}

\newcommand{\ang}{\ensuremath{\text{\AA}}}
\newcommand{\diff}{\ensuremath{\text{d}}}
\newcommand{\imp}{\rm{imp}}

\newcommand{\hc}{\rm{H.c.}}

\newcommand{\bj}{\mathbf{j}}
\newcommand{\bk}{\mathbf{k}}

\newcommand{\beq}{\begin{eqnarray}}
\newcommand{\eeq}{\end{eqnarray}}

\renewcommand{\Im}{{\rm Im}\,}
\newcommand{\hpr}[2]{\ensuremath{\langle #1|#2\rangle}}

\begin{document}

\title{Orbitally controlled Kondo effect of Co ad-atoms on graphene}

\author{T. O. Wehling}
\affiliation{1. Institut f\"{u}r Theoretische Physik, Universit\"{a}t Hamburg, D-20355 Hamburg, Germany}
\author{A. V.  Balatsky}
\affiliation{Theoretical Division, Los Alamos National Laboratory, Los Alamos, New Mexico 87545,USA}
\affiliation{Center for Integrated Nanotechnologies, Los Alamos National Laboratory, Los Alamos, New Mexico
87545,USA}
\author{M. I. Katsnelson}
\affiliation{Institute for Molecules and Materials, Radboud University Nijmegen, NL-6525 AJ Nijmegen, The
Netherlands}
\author{A. I. Lichtenstein}
\affiliation{1. Institut f\"{u}r Theoretische Physik, Universit\"{a}t Hamburg, D-20355 Hamburg, Germany}
\author{A. Rosch}
\affiliation{Institute for Theoretical Physics, University of Cologne, 50937 Cologne, Germany}
\begin{abstract}
  Based on ab-initio calculations  we identify
  possible scenarios for the Kondo effect due to Co ad-atoms on
  graphene. General symmetry arguments show that for magnetic atoms in
  high-symmetry positions, the Kondo
  effect in graphene is controlled not only by the spin but also by the
  orbital degree of freedom. For a Co atom absorbed on top of a carbon atom, the Kondo
  effect is quenched by spin-orbit coupling below an energy scale of
  $\sim \!\!15$\,K.  For Co with spin $S=1/2$ located in the center of
  a hexagon, an SU(4) Kondo model describes the entanglement of
  orbital moment and spin at higher energies, while below $\sim
  60$\,meV spin-orbit coupling leads to a more conventional SU(2)
  Kondo effect. The interplay of the orbital Co physics and the
  peculiar band-structure of graphene is directly accessible in
  Fourier transform tunneling spectroscopy or in the gate-voltage
  dependence of the Kondo temperature displaying a very strong,
  characteristic particle-hole asymmetry.
\end{abstract}
\maketitle
\section{Introduction}
Graphene differs from usual metals or semiconductors in three important aspects: It is a truly two-dimensional
material \cite{Novoselov_science2004} with the charge carriers resembling massless Dirac fermions
\cite{Geim2005,Zhang2005,RMP_AHC2009} and the chemical potential being tunable by gate voltages \cite{Novoselov_science2004}.
Recently, scanning tunneling spectroscopy experiments of graphene opened the exciting possibility to address its
electronic properties locally and to study the interaction of graphene with magnetic ad-atoms. For single Co
atoms adsorbed on heavily doped graphene, the observation of Kondo resonances with Kondo temperatures of the
order of $T_K=15K$ has been reported \cite{Mattos_09}. Theoretically, Kondo physics in ``Dirac materials'' defined as class of materials with low energy Dirac type excitations has
been first addressed in the context of high Tc superconductors
\cite{fradkin-1996,Ingersent96a,vojta0,vojta1,vojta2,RMP_Balatsky}. It has been demonstrated that even in the undoped case a Kondo
effect can exist above a certain critical coupling between the impurity spin and the Dirac electrons
\cite{fradkin-1996,Ingersent96a}. The dependence of the critical coupling, Kondo temperatures and
impurity spectral functions on doping and localized impurity states
has been studied in the context of graphene only in terms of theoretical
model systems like {\em single orbital} Anderson models or SU(2) Kondo models
\cite{Hentschel07,Sengupta07,balseiro-2009,Zhuang_STS_09,Uchoa_STS_09,Dellanna10}.

The importance of orbital physics for the Kondo effect
arises  because localized spins in magnetic ions occur almost exclusively in partially
filled $d$ or $f$ shells. For graphene the same two-dimensional representations of the hexagonal $C_{6v}$
symmetry group, that determines the orbital degeneracies of ad-atoms in high-symmetry locations, is also
responsible for the band degeneracies in graphene at the two Dirac points. Accordingly, the spin of an ad-atom
in the center of a carbon hexagon can
only couple efficiently by superexchange to graphene bands close to the Dirac point,
if it is localized in orbitally degenerate levels.
 Therefore,
the orbital degree of freedom and also spin-orbit coupling naturally govern the Kondo physics in graphene.
Indeed, recent studies \cite{Fano_STM_09,Uchoa_STS_09,saha09} showed that the tunneling into s-wave symmetric
impurity orbitals can be strongly suppressed by graphene's particular symmetries, but the decisive role of the
orbital degree of freedom has to our knowledge not been studied.
In general one can expect that correlation effects will entangle fluctuating orbital \cite{vanKempen02}
and spin degrees of freedoms. This can lead to an SU(4) Kondo effect
\cite{Borda03,DeFranceschi05}. We
show that symmetry and orbital selection rules govern not only the coupling to the graphene bands close to
the Dirac points but also to high-energy van-Hove singularities. We find that virtual high-energy fluctuations control the
size of the Kondo temperature, and, in turn, can lead to a strongly asymmetric gate-voltage dependence of $T_K$,
that would be characteristic for a specific set of orbitals.


The question, which model is appropriate to describe a certain \textit{realistic} magnetic impurity
system, is indispensable for understanding experiments as in Ref.~\onlinecite{Mattos_09} but is in general not easy:
For the classical example of Fe in Au studied since the 1930s,
an answer could be found only recently \cite{Costi_09}.
In this article, we consider the experimentally  important case of Co on graphene and develop a
\textit{first-principles} based model describing the Kondo physics in this system.
This example shows, that the non-trivial orbital structure of the impurity indeed controls the Kondo physics.

\section{Density functional simulation of Co on graphene}
For an ab-initio description of Co on graphene we performed density functional (DFT) calculations on $6\times 6$
and $4\times 4$ graphene supercells containing one Co ad-atom using the Vienna Ab Initio Simulation Package
(VASP) \cite{Kresse:PP_VASP} with the projector augmented wave (PAW) \cite{Bloechl:PAW1994,Kresse:PAW_VASP}
basis sets. To judge the role of on-site Coulomb interaction, we employed a generalized gradient
approximation (GGA) \cite{Perdew:PW91} as well as GGA+U with $U=2$\,eV, $J=0.9$\,eV and $U=4$\,eV, $J=0.9$\,eV. We obtained fully relaxed structures for all of these functionals.

\begin{figure}
 \includegraphics[width=.98\linewidth]{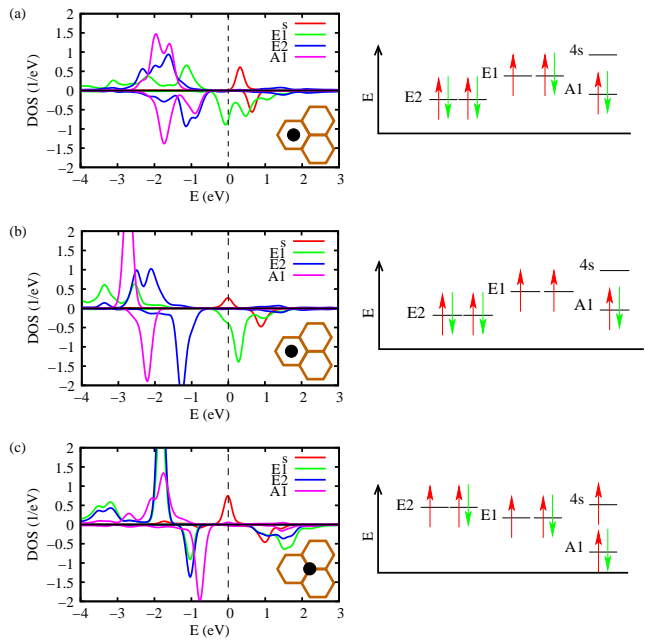}
\caption{\label{fig:LDApU_Co_LDOS}(Color online) Orbitally resolved spin-polarized local density of states (LDOS) (left) and corresponding energy level diagrams (right) for a Co
ad-atom at an h-site (a,b) and a t-site (c) for $U=2$\,eV and $J=0.9$\,eV. At the
h-site, Co has either spin 1/2 due to one hole in the 3d orbitals with E1 symmetry ($d_{xz}$, $d_{yz}$) (a) or spin 1 with two holes in the E1 orbitals (b). For Co
on a t-site we find spin 3/2 due to two holes in the Co 3d orbitals and one electron in the Co 4s
orbital.}
\end{figure}

In agreement with Refs. [\onlinecite{Kawazoe04,Mao_Co_graphene_09,Fano_STM_09}], our GGA calculations find Co positioned above the middle of a hexagon on graphene (h-site), with Co on top of carbon
(t-site) or above a bridge site (b-site) being both more than $0.5$\,eV higher in energy. GGA predicts
the electronic configuration of Co close to spin $S=1/2$ at all adsorption sites. For $U=2$\,eV, $J=0.9$\,eV
h-site adsorption is still the global total energy minimum with two different electronic configurations of Co:
First, Co can be in a spin $S=1/2$ state. The corresponding GGA+U local density of states at the Co site is depicted in Fig. \ref{fig:LDApU_Co_LDOS}a: the Co 4s-orbital is unoccupied in this configuration and one hole resides in the Co d-orbitals with E1 symmetry ($d_{xz}$, $d_{yz}$). The other solution at $U=2$\,eV, $J=0.9$\,eV (Fig. \ref{fig:LDApU_Co_LDOS}b) yields the Co d-electrons can carring approximately spin $S=1$ and a state derived from the Co s-orbital directly at the Fermi level. 
We note that this solution becomes unstable upon decreasing of $U$ and energetically unfavorable upon increasing of $U$.  At $U=2$\,eV, there exists a metastable configuration for Co at a t-site with $S=3/2$ (Fig. \ref{fig:LDApU_Co_LDOS}c), which is $0.04$\,eV higher in energy than the $S=1/2$ h-site configuration.
For $U=4$\,eV, $J=0.9$\,eV the global minimum energy is found for Co with $S=3/2$ on a
t-site, which is $0.2$\,eV and $0.08$\,eV lower in energy than the h- and b-sites, respectively.

In the following we will consider all relevant cases, first $S=3/2$ on the t-site, then $S=1$ on the h-site and, finally,
 $S=1/2$ on the h-site as the most interesting case.

\section{Scenarios for the Kondo effect}
For Co adsorbed on a t-site the crystal fields have $C_{3v}$ symmetry: the 5 d-orbitals split into
two orbital doublets corresponding to the two-dimensional representation (E) of $C_{3v}$ and a
singlet of the one-dimensional representation (A1).
As can be seen
from the LDOS (Fig. \ref{fig:LDApU_Co_LDOS} b), the spin 3/2 of the Co atom is made up by a spin 1 residing in
the Co 3d-orbitals of E symmetry ferromagnetically coupled to a spin 1/2 mainly from the Co 4s orbital.
The four low-energy graphene bands  close to the Dirac points can also be decomposed into one two-dimensional and
two one-dimensional representation which hybridize efficiently with the E and A1 orbitals, respectively.
In the absence of spin-orbit coupling (SOC) one can expect a two-stage Kondo effect: first, the direct coupling of the
s-orbital to the C atom beneath quenches 1/2 of $S=3/2$ resulting in a spin 1 coupling to the two bands
of $E$ symmetry via the next-nearest C atoms which screen the remaining spin in a second stage.
To estimate
the strength of SOC, we calculate the magnetic anisotropy,
$E_{\rm MAE}=E_{||}-E_{\perp}\approx 1.0$\,meV, as the
energy difference between magnetization parallel and perpendicular to the graphene plane in GGA+U with
$U=4.0$\,eV and $J=0.9$\,eV. This translates into different energies, $E_{|S_z|=1/2}-E_{|S_z|=3/2}\approx
1.3$\,meV, depending on the $z$-component, $S_z$, of the Co spin $S=3/2$ in this configuration. Ignoring Kondo physics,
the ground state
has $|S_z|=3/2$ and exhibits Kramers degeneracy but spin flips $S_z=3/2\rightarrow S_z=-3/2$ induced by
electron scattering are only possible in higher order processes.
Hence, the Kondo effect is efficiently suppressed for Co in this configuration,
as soon as $T_K \ll 1.3$\,meV$\approx 15$\,K. For a (first stage) Kondo temperature of the order of $15$\,K
or above, a definite determination of the relevant low-energy model is very difficult, but a possible scenario is
that first the Kondo effect partially screens $1/2$ out of $S=3/2$, then spin-orbit coupling stabilizes a low
energy doublet  ($S_z=\pm 1$), which is finally screened at very low temperatures by an anisotropic Kondo effect.

For Co at an h-site, the situation is more interesting and a quantitative analysis easier: 
crystal fields of
the $C_{6v}$ symmetry decompose the d-orbitals into two non-equivalent two-dimensional irreducible
representations E1 ($d_{xz}$, $d_{yz}$) and E2 ($d_{x^2-y^2}$, $d_{xy}$) plus one one-dimensional 
representation A1 ($d_{3z^2-r^2}$). For both, the $S=1/2$ and the $S=1$ configuration, the spin mainly resides in the E1 orbitals, as shown in Fig. \ref{fig:LDApU_Co_LDOS} a) and b). We calculated the strength of the SOC for Co on graphene using VASP and
obtained $\lambda=60$\,meV for the single particle SOC operator $H_{\rm SOC}\approx \lambda
\mathbf{l}\cdot\mathbf{s}$, with $\mathbf{l}$ and $\mathbf{s}$ being the orbital and spin angular momentum
operators, respectively. Moreover, in the GGA
calculations we obtained the crystal
field splitting from the d-level energies as $\epsilon_{E2}-\epsilon_{E1}=-0.8$\,eV
and $\epsilon_{A1}-\epsilon_{E1}=-0.56$\,eV for Co at the
h-site.
 Diagonalizing a Co atom with $S=1$ in $d^8$ configuration in this crystal field yields a singlet as the
ground state, which is separated by about $0.008$\,eV$\approx 90$\,K from a doublet of first
excited states. Hence, for a high-spin Co at an h-position an $S=1$ Kondo effect is
quenched if $T_K\lesssim 90$\,K and a much lower $T_K$ \cite{Mattos_09} is only consistent with the
 low-spin configuration.

For Co at an h-site with $S=1/2$ in $d^9$ configuration (see Fig. \ref{fig:LDApU_Co_LDOS} a) one obtains a four-fold degenerate state. SOC lifts
this degeneracy, resulting in a twice degenerate atomic ground state, which is separated from a doublet of excited
states by an energy of the order of $\lambda$. In this more than half-filled regime, the spin- and orbital
moment are aligned in parallel (c.f. Hund's 3rd rule). The d-hole resides in the highest
crystal field orbitals, $E1$, which have  $|l_z|=1$. Hence, the Zeeman splitting for out-of-plane magnetic
fields, $B_z$, is $\Delta E=\mu_B B_z(g_l s_z+g_l l_z)/\hbar=\pm\mu_B 2B_z$ resulting in the effective g-factor
of $g |s_z|/\hbar=2$. The SOC induced lifting from four fold to two fold degeneracy will
lead to SU(4) Kondo physics above the scale of $\lambda$ and SU(2) Kondo physics at
lower energies.

\section{Kondo effect of Co at an h-site}
To address the Kondo effect in this configuration, we describe the Co at an h-site in terms of an
Anderson impurity model: The conduction electrons residing in graphene's $\pi$-bands are modeled by a
tight-binding Hamiltonian
with $t=-2.97$\,eV, $t'=-0.073$\,eV, and $t''=-0.33$\,eV quantifying the nearest, next-nearest 
and next-to-next nearest neighbor hopping, respectively \cite{Reich:2002}.
For the Co atom, we consider its 3d orbitals,
$\hat{H}_{\imp}=\sum_{m,\sigma}\epsilon_{|m|} n_{m,\sigma}+\frac{U}{2}\sum_{(m,\sigma)\neq (m',\sigma')}n_{m,\sigma}
n_{m',\sigma'}$ with $n_{m,\sigma}=d^\dagger_{m,\sigma}d_{m,\sigma}$, where $m$ is the quantum number of
the z-component of the orbital momentum, $d_{m,\sigma}$ are Fermi operators,  $U$ is the local Coulomb repulsion
and $\epsilon_{|m|}$ are the bare on-site energies. Here, we include hopping from the localized d-orbital to the nearest-neighbor C-atoms and use the $C_{6v}$ symmetry
to write the coupling of Co to graphene in the form
\begin{equation}
\hat{V}=\sum_{m,\sigma}V_{|m|}c^\dagger_{m,\sigma}d_{m,\sigma}+\hc,
\end{equation} where $c_{m,\sigma}=\sum_{<\bj>}e^{im\phi_\bj}c_{\bj,\sigma}/\sqrt{6}$, $c_{\bj,\sigma}$ is the Fermi operator of electrons at carbon atom at site $\bj$ and
$\phi_\bj$ is the angle between a fixed crystalline axis and the bond from site $\bj$ to the Co impurity. All local physics, is contained in the local Hamiltonian, $H_{\imp}$,
and the hybridization function $\Delta_{mm'}(\omega)$ defined as
$\Delta_{mm'}(i\omega)=V_{|m|}G^0_{mm'}(i\omega)V_{|m'|}$, where
\begin{equation}
 G^0_{mm'}(i\omega)=\int\diff k\hpr{m}{\bk}(i\omega-H_k)^{-1}\hpr{\bk}{m'}
\label{eqn:G0_E1_E2}
\end{equation}
is the bare graphene electron Green function of the states $c_m$. $\Delta_{mm'}(i\omega)=\Delta_{m}(i\omega)\delta_{mm'}$ is diagonal and $\Delta_{m}(i\omega)=\Delta_{-m}(i\omega)$ by symmetry.

The hybridization functions for different values of $|m|=0,1,2$ are subject to selection rules imposed by the
matrix elements $\hpr{m}{\bk}$: The eigenstates of $\hat{H}_0$ close to the Dirac points, $K$ and $K'=-K$,
transform according to E1 and E2 under $C_{6v}$ with the E1 and E2 being degenerate at the Dirac point. Hence,
hybridization with $m=0$ states is cubically suppressed and
$\Im\,\Delta_{|m|=1}(\omega)/V^2_1=\Im\,\Delta_{|m|=2}(\omega)/V_2^2\approx
-\pi\frac{\sqrt{3}|\omega|}{2\pi(t-2t'')^2}$ to leading order in $\omega$.

In contrast to the particle hole symmetry for $\omega \to 0$,
the hybridization
functions are largely asymmetric at higher energies. This is caused by the E1 and the E2 impurity orbitals coupling
each to only one of the van-Hove singularities resulting from the graphene bands at the Brillouin zone M point:
The E1 impurity orbitals as well as the graphene valence electron wave functions at the M point are odd
under 180$^\circ$ rotation about the h-site, whereas the E2 orbitals and conduction electron wave functions at
the M point are even under this transformation. Hence, the E1 hybridization exhibits a logarithmic singularity,
$\Delta_1(\omega)\sim\ln|\omega-E_{M-}|$, at $E_{M-}=t + t' - 3 t''\approx - 2.1$\,eV. However, there is no singularity in the E1 hybridization
at the energy of the conduction band van-Hove singularity, $E_{M+}=-t + t' +3 t''\approx 1.9$\,eV. For the E2
orbitals, the situation is reversed: $\Delta_2(\omega)\sim\ln|\omega-E_{M+}|$ for
$\omega\rightarrow E_{M+}$.

\begin{figure}
\includegraphics[width=.8\linewidth]{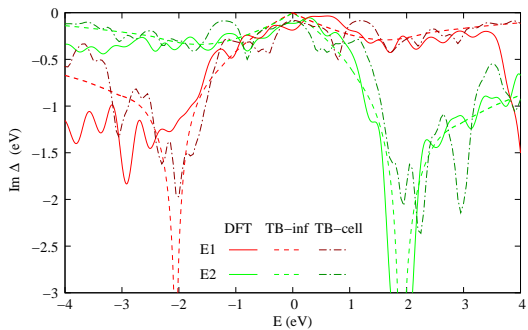}
\caption{\label{fig:Delta_fit_NNN} (color online) Imaginary part, $\Im\,\Delta_{|m|}(E)$, of the hybridization
functions of the E1 ($|m|=1$) and E2 orbitals ($|m|=2$) of a Co ad-atom adsorbed to a graphene h-site.
Hybridization functions obtained from DFT and tight-binding (TB) models of one Co on an infinite graphene sheet (TB-inf) as well as the same supercell (TB-cell) as used in DFT with $V_1=1.4$\,eV and $V_2=1.5$\,eV
are shown.}
\end{figure}

To obtain realistic hybridization strengths, $V_1$ and $V_2$, we calculate $\Delta_{m}$ by means of DFT as
described in Ref. \onlinecite{Fano_STM_09} and fit the tight-binding hybridizations via $V_1$ and $V_2$ (see Fig.
\ref{fig:Delta_fit_NNN}).
The tight-binding (TB) hybridizations are obtained in two ways: (1) by directly evaluating Eq. (\ref{eqn:G0_E1_E2}), which models one Co ad-atom on an infinite graphene sheet (TB-inf), and (2) by employing the same $6\times 6$ supercell as in the DFT calculations and performing the same supercell Brillouin zone integration (TB-cell).

The high energy particle-hole / E1-E2 asymmetry is striking the DFT as well as in both TB hybridization functions. The DFT hybridization functions display small wiggles and the van-Hove singularities appear to be smeared out. Comparison of the DFT hybridization to the TB supercell hybridization shows that these two effects are supercell artifacts. The tight-binding curves can be well fitted to DFT\footnote{Note that only \textit{one} fitting parameter is employed per curve. Including an on-site potential at the adjacent C-atoms allows bringing the energy positions of the wiggles in the TB supercell and the DFT hybridization functions into agreement. Here, we focus on qualitative consequences of the particle hole asymmetry in the hybridization function and do not include a second fitting parameter.} with $V_1=1.4$\,eV and $V_2=1.5$\,eV. For energies above $3.4$\,eV and below $-3.1$\,eV also further bands contribute to $\Delta_m(\omega)$. They contribute to screening and
lead to a (finite) renormalization of the exchange coupling which we absorb in a redefinition of the bare exchange coupling $J_0$ used below. In the following, we employ the tight-binding hybridization function for one Co on an infinite graphene sheet, which is free from supercell artifacts.

To estimate Kondo temperatures and their gate voltage dependence, we solve the scaling equation
\cite{HewsonBook,kuramoto}
\begin{equation}
\label{eqn:scaling}
 \frac{\diff J(D)}{\diff D}=-N(D)J^2(D) \frac{\rho(\mu-D)+\rho(\mu+D)}{2D},
\end{equation}
where $J(D)$ is the renormalized exchange coupling, $D$ the high energy cut-off, $\mu$ the chemical potential in
graphene and $\rho(\omega)=-\Im\Delta_1(\omega)/(\pi V^2_1)$. The degeneracy factor, $N(D)=4$ for $D>\lambda$
and $N(D)=2$ else \cite{kuramoto}, accounts for locking the orbital- to the spin-degree of freedom below the energy scale of the
spin-orbit coupling. 

\begin{figure}
 \includegraphics[width=0.48\linewidth]{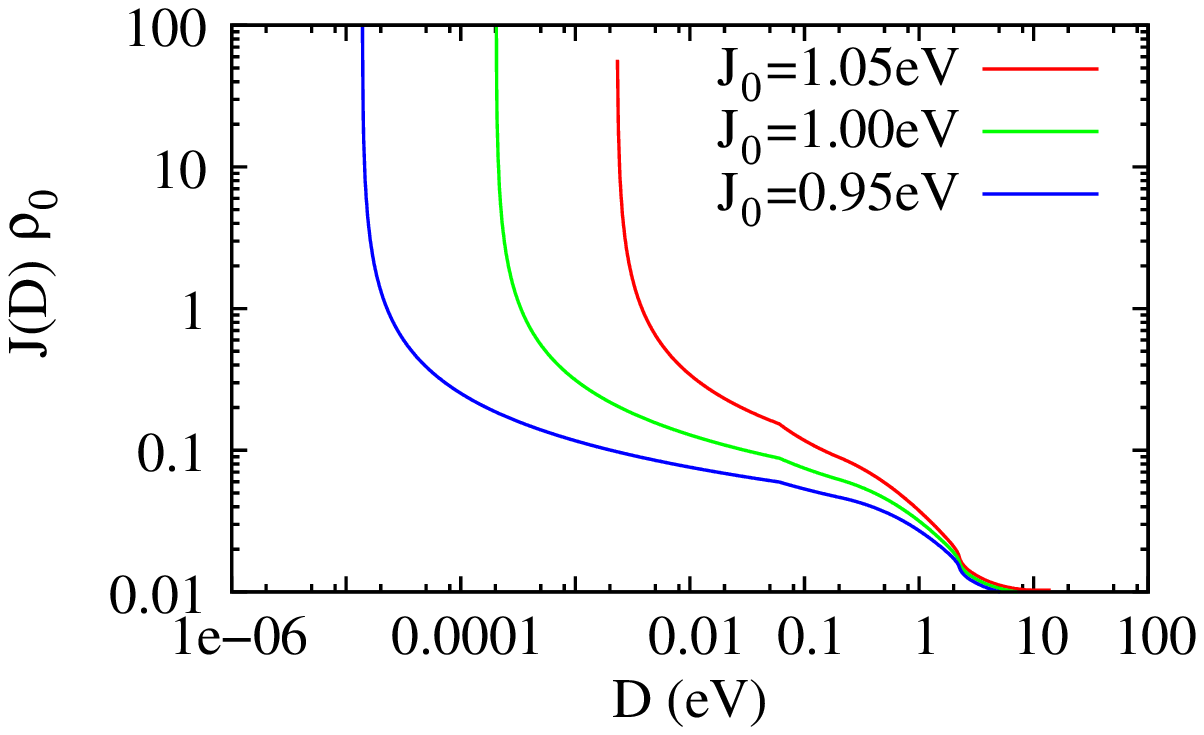}
\includegraphics[width=0.48\linewidth]{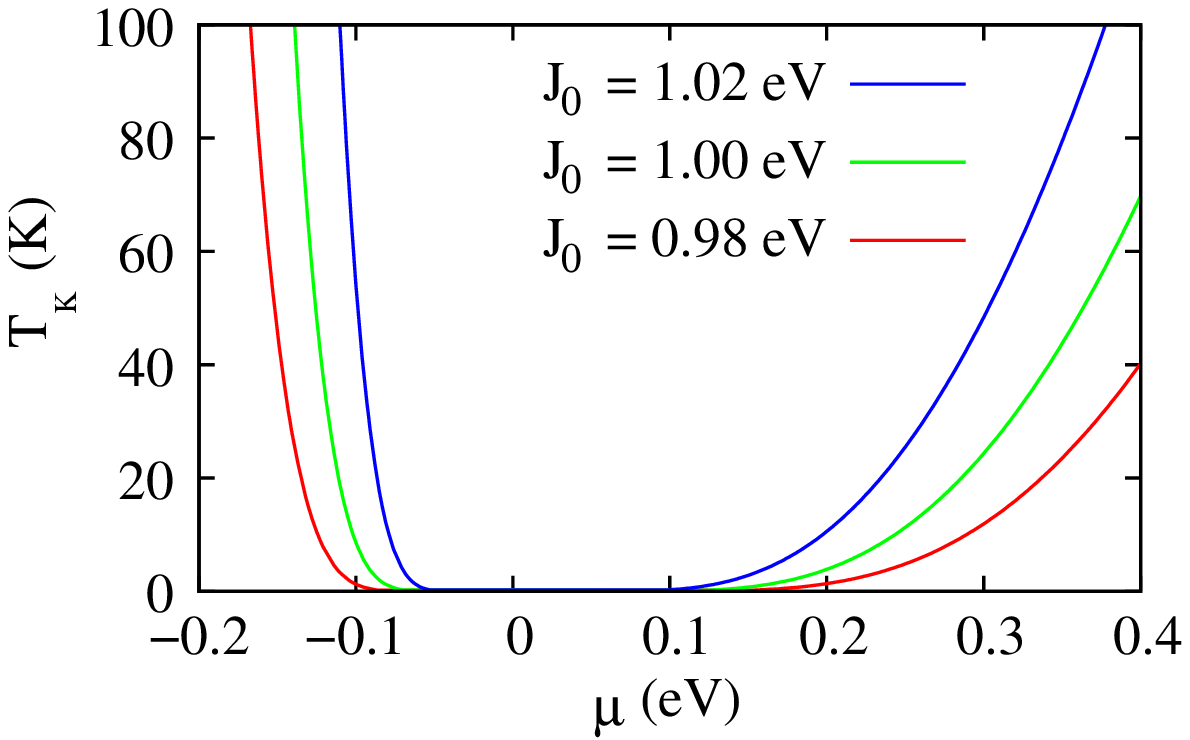}
\caption{\label{fig:Jflow_Tk} (color online) Left: Renormalized effective coupling strength,
$J(D)\frac{\rho(\mu-D)+\rho(\mu+D)}{2}\equiv J(D)\rho_0$, as function of the high energy cut-off $D$ for the
chemical potential $\mu=0.20$\,eV and three different bare couplings $J_0=1.05$, $1.00$, and $0.95$\,eV.  Right:
Kondo temperatures $T_K$ as function of $\mu$ for $J_0=1.02$, $1.00$, and $0.98$\,eV. The Kondo temperatures are
remarkably asymmetric for electron and hole doped graphene.}
\end{figure}

Like all one-loop renormalization group equations, Eq. (\ref{eqn:scaling}) is valid as long as the renormalized coupling is small, see Fig. \ref{fig:Jflow_Tk}, left. It is used
to  detect the energy scale where the strong coupling regime is approached which we identify with the Kondo temperature. This procedure correctly identifies the exponentially strong sensitivity of the Kondo scale on system parameters. While an exact calculation of the prefactor of $T_K$ in the limit of small $J_0$ and $\mu \neq 0$ requires at least a two-loop calculation\cite{HewsonBook}, the one-loop equation (\ref{eqn:scaling}) captures the main effect of a frequency-dependent density of states. Note, however, that the
 perturbative renormalization group calculation cannot describe the quantum-critical point\cite{fradkin-1996,Ingersent96a,vojta0,vojta1,vojta2} obtained for vanishing density of states as the renormalized coupling is not small in this case.

Varying $J_0$ for $\mu=0.2$\,eV (as in the experiment reported in Ref.\cite{Mattos_09}), we find the Kondo temperature
changing by an order of magnitude for varying $J_0$ within a few percent. (See Fig. \ref{fig:Jflow_Tk} left.)
While this hinders predictions of the absolute value of the Kondo temperature, the trend of how
the Kondo temperature depends
on the chemical potential in graphene is robust w.r.t. changes in $J_0$: Leaving $J_0$ as fitting parameter, we
predict the gate voltage-dependence of $T_K$ as shown in Fig. \ref{fig:Jflow_Tk} right.
Note that the used values for $J_0\approx 1$eV are of order $V_1^2/U$ for realistic values of $V_1$ and $U$.
The remarkable asymmetry of the hybridization function leads to a highly asymmetric dependence of the Kondo
temperatures on the gate voltages. If the E2 orbitals were carrying the magnetic moment instead of E1 this
asymmetry would be reversed.

Interestingly, for $J_0>J_c\approx 1.1$\,eV we find that the Kondo effect persists even
for vanishing doping. This implies that
by relative small changes (e.g. using different substrates) it may be possible to realize the quantum critical point
of the pseudogap Kondo problem \cite{fradkin-1996,Ingersent96a,RMP_Balatsky,vojta0,vojta1,vojta2}.


\section{Fourier transformed STM}
The symmetry of Co orbital carrying the magnetic moment can be probed by FT-STS. In the simplest model (see e.g. Ref. \onlinecite{bena:2005}), FT-STS measures the Fourier transform of the local density of states, $|\rho_k(E)|$, in the vicinity of an impurity with the constant background of a clean sample being subtracted: 
\begin{equation}
\rho_k(E)=-\frac{1}{\pi}\int\diff^2 r e^{ikr} \Im\;(G(r,r,E)-G^0(r,r,E)).
\label{eqn:FT-STS_def}
\end{equation}
Here, $G(r,r,E)$ denotes the full Green function of the graphene-impurity system in position space representation and $G^0(r,r,E)$ is the Green function of clean graphene.
Using a resonant level model for the Kondo peak, we employ the  T-matrix formalism (see e.g. Refs.
\onlinecite{bena:2005,RMP_Balatsky,imp_loc-2007-}) in the discrete position space representation. Then, Eq. (\ref{eqn:FT-STS_def}) leads to
\begin{widetext}
 \begin{equation}
\label{eqn:FT-STS2}
\rho_k(E)=-\frac{1}{\pi}\sum_{j}\int\diff^2 k' e^{ikr_j}\frac{1}{i}\left[\delta G_{k',k'+k}(E)-\delta G^*_{k'+k,k'}(E)\right]_{jj},
\end{equation}
\end{widetext}
where the index $j$ labels the two atoms per graphene unit cell, $r_j$ their position w.r.t. the unit cell origin and the $k'$ integral extends over the first Brillouin zone. The Green functions occurring in Eq. (\ref{eqn:FT-STS2}) are $2\times 2$ matrices in sublattice space and obtained from the unperturbed graphene Green functions $G^0_k(E)$ by using the $T$-matrix: $\delta G_{k',k'+k}(E)=G^0_{k'}(E)T_{k',k'+k}(E)G^0_{k'+k}(E)$.

In a resonant level model for the Kondo peak, we consider orbitals of E1 and E2 symmetry to derive the FT-STS patterns from Eq. (\ref{eqn:FT-STS2}) using a corresponding $T$-matrix with phase $\pi/2$. 

\begin{figure}
\includegraphics[width=.98\linewidth]{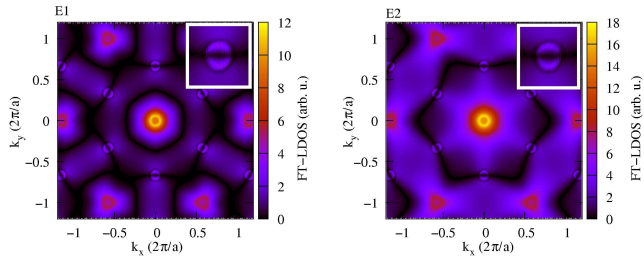}
\caption{\label{fig:FT-STS_E1_E2}(color online) Fourier transformed LDOS for
resonances of E1 (left) and E2 symmetry (right) at energy $E=0.4$\,eV
for a phase  $\pi/2$ of the t-matrix at this energy.
The reciprocal
lattice vectors are $(2/\sqrt{3},0)$ and $(1/\sqrt{3},1)$ in units of $2\pi/a$, where $a=2.465\ang$ is the
lattice constant. The bright spots close to the panel boundaries are centers of higher Brillouin zones. The insets show close-up views of the K/K' points which are at symmetry equivalent positions of $(0,2/3)$.}
\end{figure}

The resulting Fourier transformed LDOS images are shown in Fig. \ref{fig:FT-STS_E1_E2}. As the Kondo
impurity on the h-site couples equally strong to both K and K' points, the inter-valley
scattering is very strong. Due to the two sublattices, it depends however
strongly on the energy $E$ and the phase shift $\delta$ to which extent
this K-K' scattering leads to FT-STS intensity at the K and K' points. As Fig. \ref{fig:FT-STS_E1_E2} shows,
there is a double arc structure of intensity around K / K' for $E=0.4$\,eV with the radius
given by twice the Fermi wave vector. These structures disappear for $E\rightarrow 0$.
The orbital symmetries manifest in distinct FT-STS maps with characteristic gate voltage and tunneling bias dependence: Upon energy, $E\rightarrow -E$, and phase shift reversal, $\pi/2+\eta\rightarrow\pi/2-\eta$, the FT-STS patterns of resonances with E2 and E1 symmetry interchange.

\section{Conclusions}
We showed that the Kondo effect of Co ad-atoms on graphene is
controlled by the particular symmetries of the Co 3d orbitals
originating from graphene crystal field splitting. Based on
first-principles calculations we found different possible scenarios
with $t-$ or $h$-site adsorption of Co and consequences for the Kondo
physics. For Co at an $h$-site we found a surprising asymmetry of
Kondo temperatures w.r.t. the chemical potential and predicted
characteristic FT-STS patterns. Both of these effects can be probed by
STM. 

The importance of the orbital degree of freedom for the Kondo effect
in graphene can be traced back to the symmetries underlying the
peculiar band degeneracies of graphene at the Dirac point. Therefore
the orbital degree of freedom is expected to control the Kondo physics
in graphene also for other magnetic impurities occupying high-symmetry
positions.

{\em Acknowledgments}. We are grateful to R. Bulla, J. von Delft, M. Vojta and especially H. Manoharan for useful discussions.
We thank I. Kolorenc for providing us his
exact diagonalization code. This work was supported by Stichting voor Fundamenteel Onderzoek der Materie (FOM),
the Netherlands, by SFB-668(A3), SFB 608 and SFB TR12 of the DFG, the Cluster of Excellence ``Nanospintronics'' (LExI Hamburg), and Grant No. PHY05-51164 of the NSF. Work At Los Alamos was performed at the Center for Integrated Nanotechnologies, a U.S. Department of Energy, OBES (Contract DE-AC52-06NA25396). Computer time at HLRN is acknowledged.

\bibliography{graphene_s}

\end{document}